\documentclass[]{spie}  

\usepackage{amsmath,amsfonts,amssymb}
\usepackage{wrapfig,booktabs}
\usepackage{graphicx}
\usepackage{wrapfig}
\usepackage{makecell}
\usepackage{xcolor}
\usepackage{textgreek}
\usepackage[colorlinks=true, allcolors=blue]{hyperref}

\usepackage[normalem]{ulem}
\usepackage{color}

\title{Developing a New Generation of Integrated Micro-Spec Far Infrared Spectrometers for the EXperiment for Cryogenic Large-Aperture Intensity Mapping (EXCLAIM)}
\author[a]{Carolyn G. Volpert}
\author[b]{Emily M. Barrentine}
\author[b]{Mona Mirzaei}
\author[b]{Alyssa Barlis}
\author[a]{Alberto D. Bolatto}
\author[b]{Berhanu Bulcha}
\author[b]{Giuseppe Cataldo}
\author[b,c]{Jake A. Connors}
\author[b]{Nicholas Costen}
\author[b]{Negar Ehsan}
\author[b]{Thomas Essinger-Hileman}
\author[b]{Jason Glenn}
\author[b]{James P. Hays-Wehle}
\author[b]{Larry A. Hess}
\author[b]{Alan J. Kogut}
\author[d]{Harvey Moseley}
\author[e]{Jonas Mugge-Durum}
\author[b]{Omid Noroozian}
\author[f]{Trevor M. Oxholm}
\author[b]{Maryam Rahmani}
\author[b]{Thomas Stevenson}
\author[b]{Eric R. Switzer}
\author[b]{Joseph Watson}
\author[b]{Edward J. Wollack}

\affil[a]{University of Maryland, College Park, MD, USA}
\affil[b]{NASA Goddard Space Flight Center, Greenbelt, MD, USA}
\affil[c]{National Institute of Standards and Technology, Boulder, CO, USA}
\affil[d]{Quantum Circuit, Inc., USA}
\affil[e]{Deutsches Zentrum
für Luft und Raumfahrt, Köln, DE}
\affil[f]{University of Wisconsin-Madison, Madison, WI, USA}

\authorinfo{Further author information: (Send correspondence to C. Volpert)\\C. Volpert: E-mail: cvolpert@astro.umd.edu}
\begin{document} 
\maketitle

\begin{abstract}
The current state of far-infrared astronomy drives the need to develop compact, sensitive spectrometers for future space and ground-based instruments. Here we present details of the $\rm \mu$-Spec spectrometers currently in development for the far-infrared balloon mission EXCLAIM. The spectrometers are designed to cover the 555 – 714\,$\rm \mu$m range with a resolution of R = $\rm \lambda / \Delta\lambda$ = 512 at the 638\,$\rm \mu$m band center. The spectrometer design incorporates a Rowland grating spectrometer implemented in a parallel plate waveguide on a low-loss single-crystal Si chip, employing Nb microstrip planar transmission lines and thin-film Al kinetic inductance detectors (KIDs). The EXCLAIM $\rm \mu$-Spec design is an advancement upon a successful R = 64\,$\rm \mu$-Spec prototype, and can be considered a sub-mm superconducting photonic integrated circuit (PIC) that combines spectral dispersion and detection. The design operates in a single $M{=}2$ grating order, allowing one spectrometer to cover the full EXCLAIM band without requiring a multi-order focal plane. The EXCLAIM instrument will fly six spectrometers, which are fabricated on a single 150\,mm diameter Si wafer. Fabrication involves a flip-wafer-bonding process with patterning of the superconducting layers on both sides of the Si dielectric. The spectrometers are designed to operate at 100 mK, and will include 355 Al KID detectors targeting a goal of NEP ${\sim}8\times10^{-19}$ W/$\rm \sqrt{Hz}$. We summarize the design, fabrication, and ongoing development of these $\rm \mu$-Spec spectrometers for EXCLAIM.
\end{abstract}


\section{Spectrometer Overview}
\label{sec:overview}

The spectrometers we describe are designed for the Experiment for Cryogenic Large-Aperture Intensity Mapping mission (EXCLAIM), a balloon-borne instrument designed to perform far-infrared spectroscopy in the $ 555{–}714\ \mu$m ($420{-}540$\,GHz) range with resolution $R{=}512$ in the band center \cite{EXCLAIMjatis}. EXCLAIM employs six integrated spectrometers that continue to use the $\mu$-Spec architecture, operating at $M{=}2$ order. EXCLAIM will fly as a conventional balloon mission with a duration ${<}24$\,hr from NASA's Ft. Sumner, New Mexico facility, and the first flight scheduled for Fall 2023.  The instrument has an aperture of $0.9$\,m, resulting in 4'\,FWHM beam, and is fully cryogenic, with the optics operating at $1.7$\,K and the spectrometers at $100$\,mK.





\subsection{Spectrometer Design}
\label{sec:specdesign}

\begin{table}[ht]
\caption{EXCLAIM's spectrometer parameters (from design and simulated performance)} 
\label{tab:fonts}
\begin{center}       
\begin{tabular}{lc} 
\hline
\hline
\rule[-1ex]{0pt}{3.5ex}  \textbf{Number of spectrometers} & 6  \\
\hline
\rule[-1ex]{0pt}{3.5ex}  \textbf{Spectrometer spectral band} & $555{–}714$\,$\rm \mu$m ($420{-}540$\,GHz)  \\
\hline
\rule[-1ex]{0pt}{3.5ex}  \textbf{Spectrometer grating order, M} & 2 (single order)   \\
\hline
\rule[-1ex]{0pt}{3.5ex}  \textbf{Spectrometer resolving power, R} & \makecell{512 at 472 GHz (center frequency) \\ 535 to 505 over spectral band}
\\
\hline
\rule[-1ex]{0pt}{3.5ex}  \textbf{Spectrometer efficiency} & 24\%  \\
\hline
\rule[-1ex]{0pt}{3.5ex}  \textbf{KID NEP (at input to each KID)} & \makecell{8 × 10$^{-19}$ W/$\rm \sqrt{Hz}$ at 0.16 fW (at KID)\\ at 5 to 26 Hz acoustic frequency}  \\
\hline
\rule[-1ex]{0pt}{3.5ex}  \textbf{Number of receivers/KIDs per spectrometer} & 355  \\
\hline
\rule[-1ex]{0pt}{3.5ex}  \textbf{KID readout band} & 3.25 to 3.75 GHz \\
\hline
\rule[-1ex]{0pt}{3.5ex}  \textbf{Operating temperature} & 100 mK \\
\hline
\hline

\end{tabular}
\end{center}
\end{table} 

\begin{figure} [ht]
   \begin{center}
   \begin{tabular}{c} 
   \includegraphics[height=10cm]{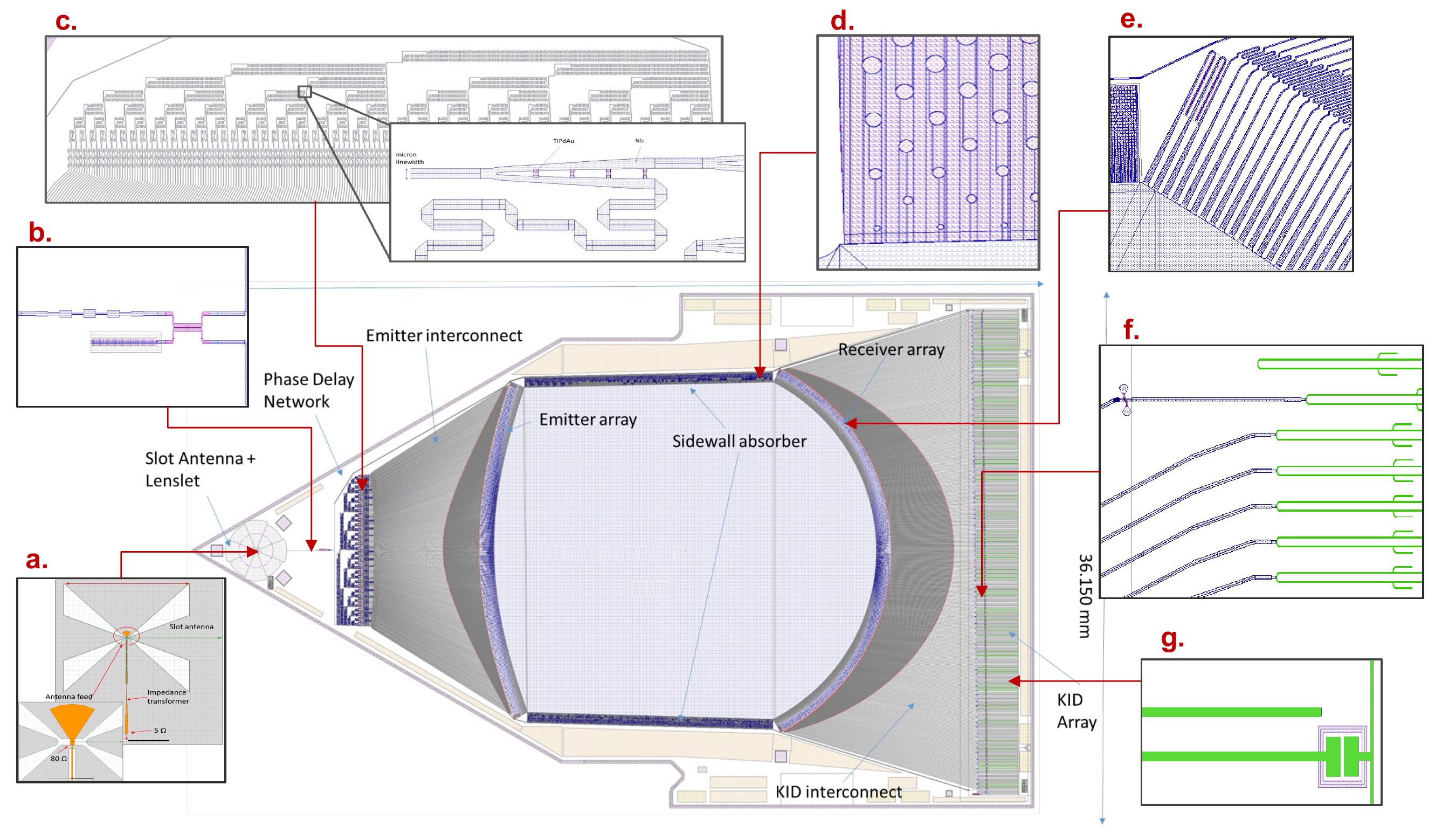}
   \end{tabular}
   \end{center}
   \caption[example] 
   {\label{fig:cutouts} A labeled schematic of an EXCLAIM spectrometer with detailed cutouts of individual components: \textbf{a)} The spectrometer's x-slot antenna depicted with the impedance transformer and a secondary transformer, shown towards the bottom of the image. The cutout on the bottom left shows a closeup of the antenna feed. \textbf{b)} An order-choosing filter between the antenna and the phase-delay network passes all even orders, and a combination of quasi-optical filters before the spectrometer's antenna and the inherent cutoff imposed by the Nb feedline gap energy result in the overall selection of a single order: $M=2$. On the right, the main feedline is coupled to a reference KID, which measures total power. \textbf{c)} The spectrometer's phase-delay network. A cutout shows a closeup example of some of the meandered Nb transmission lines, which vary in path length. \textbf{d)} A closeup of the sidewall of the 2D interference region, designed to terminate the portion of the light reflected by the receiver array and input radiation outside the EXCLAIM wavelength band. \textbf{e}) A closeup of the top portion of the receiver array. The receiver horns feed into Nb transmission lines that connect to the spectrometer detector array. \textbf{f)} The Nb lines from the receiver array connect to an array of thin-film Al microstrip KIDs. Detectors designed to diagnose noise and stray light, such as those uncoupled from the rest of the spectrometer ('dark' detectors) and those whose path skips the interference region, are included at the edges of the KID array. \textbf{g)} The KIDs in the array are all coupled via parallel-plate capacitors to the readout transmission line.
   }
\end{figure}

After light travels through the telescope's cold optics, it is focused onto a focal plane that contains six 4\,mm-diameter hyper-hemispherical silicon lenslets, each attached to the back of a spectrometer chip (Figure \ref{fig:package}). 
Each lenslet has a parylene-C anti-reflection (AR) coating 126\,$\rm\mu$m in thickness to provide optimized coupling from free-space to the Si lenslet over a curved surface for the EXCLAIM band. The light incident on the lenslet is then coupled into each spectrometer layer on the opposite side of the wafer (Figure \ref{fig:sideview}) through an x-slot antenna (Figure \ref{fig:cutouts}a), where an impedance transformer then modifies the signal to match the impedance of the spectrometer's Nb microstrip transmission lines, which are of microstrip design. The light then travels through an order-selecting filter that selects the $M{=}2$ order (Figure \ref{fig:cutouts}b) while rejecting the $M{=}1$ and $M{=}3$ orders. Higher orders are cut off by the Nb transmission line superconducting energy gap. The light then travels into a network of Nb transmission lines of varying meandered lengths that impose an approximately linear phase gradient across the array with corrections for dispersion (Figure \ref{fig:cutouts}c).Corrections for dispersion are accounted for in the single-order design, accommodation of multiple orders complicates the ability to correct for dispersion, motivating the EXCLAIM spectrometers' single-order design. The light then is sent to an array of Nb microstrip feed structure emitters, and emitted into a 2D interference region consisting of a parallel-plate waveguide formed by the top and ground plane Nb superconducting layers and the single-crystal Si dielectric. Across the 2D region, the light interferes constructively and destructively as a product of the phase delay, analogous to the operation of a classic free-space spectrometer grating.


\begin{wraptable}{r}{10cm}
    \label{wrap-tab:1}
    \begin{tabular}{l|c}
    \textbf{Efficiency Contributions} & \textbf{Design Estimate
    }\\\toprule
    Slot Antenna Coupling & 53 $\pm$ 5.5\%\\  \midrule
    Order-Choosing Filter Transmission &$\gtrsim$ 98.7\%\\ \midrule
    Reference coupler & 96.84\%\\  \midrule
    Rowland Circle Receiver Coupling &$\sim$ 50\%\\  \midrule
    MKID Coupling &$>$ 99\%\\  \midrule
    Transmission Line Loss &$\gtrsim 
    94\%$\\  \midrule
    \textbf{Total Efficiency Estimate} &\textbf{$\gtrsim$ 24\%}\\  \bottomrule
    \end{tabular}
    \vspace{1pt}
    \caption{The estimated efficiencies of the spectrometer components.}
\end{wraptable}

The light is then coupled into an array of microstrip feed receiver structures on the far side of the interference region (Figure \ref{fig:cutouts}e). Any light reflected into the interference region is terminated by metamaterial sidewall absorbers (Figure \ref{fig:cutouts}d). The 355 receiver structures are spaced to meet the Nyquist-sampling criteria while optimizing coupling efficiency and isolation. The light then travels from the receiver array through transmission lines to a corresponding array of KIDs (Figure \ref{fig:cutouts}e). Further details of this spectrometer Rowland architecture can be found in Ref.~\cite{CATALDO2019155}. The detector array contains 355 KIDs, each consisting of two branches of a half-wave microstrip transmission-line resonator, with a tuning stub feature implemented to enable post-optical-measurement correction of resonance frequencies, which can vary from design values due to fabrication tolerances (Figure \ref{fig:cutouts}f). The Nb transmission line that serves as an optical input for the detectors lies between the two symmetric half-wave resonator branches, providing isolation between the microwave signal and sub-millimeter optical signals. On the other side, one end of each resonator is coupled to the readout transmission line through a parallel-plate coupling capacitor, with each resonator designed to have a coupling $\rm Q_c \sim 2\times10^4$ (Figure \ref{fig:cutouts}g). Since each detector is tuned to a different resonant frequency, the detector array can all be read out along one transmission line. There are five additional 'reference' resonators on the spectrometer that are not coupled to the optical signal, and are instead read out as references for device performance. By design, the largest spacing between resonant frequencies will be $1.357$\,MHz, and the spectrometer will have a total bandwidth of $490$\,MHz, with a $2.678$\,MHz gap in the band center to avoid the local oscillator tone. 

\begin{center}
   \begin{figure} [ht]
   \includegraphics[height=6.5cm]{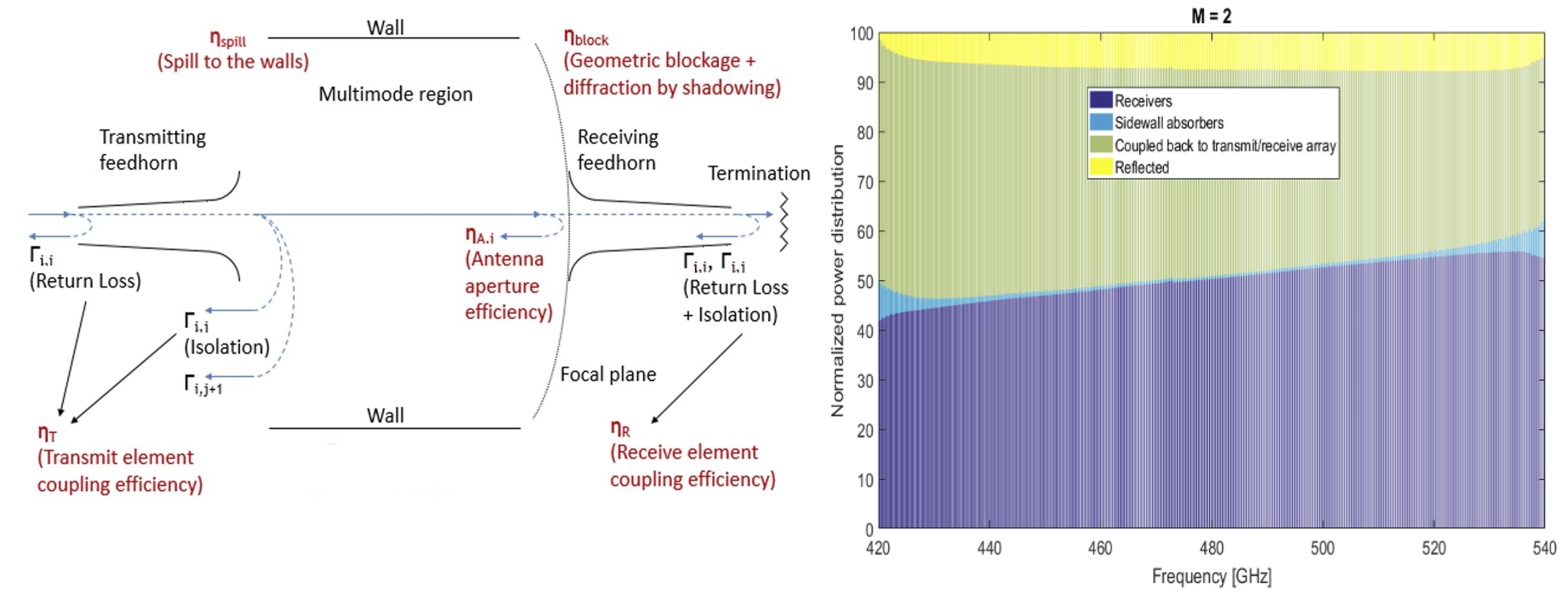}
   \caption[example] 
   { \label{fig:rec_eff} 
    Left: A diagram of the spectrometer elements relevant to calculating the Rowland circle receiver array efficiency. The transmitting feedhorn on the left corresponds to the parts of the 'Emitter array' region labeled in Figure \ref{fig:cutouts}, and the receiving feedhorn corresponds to units of the 'Receiver array' region labeled in Figure \ref{fig:cutouts}. Right: A simulation demonstrating the distribution of power as light travels through this portion of the spectrometer. The coupling efficiency of the transmitting element, the coupling efficiency of the receiving element, and the aperture efficiency of the receiver antenna all factor into the amount of optical power that is coupled into the detector array. Spill onto the sidewall increases at the red and blue edges of the optical response, and the return loss of the transmitting element introduces an additional loss of optical power due to reflections. The rest of the optical power is transmitted successfully through the region.}
   \end{figure} 
\end{center}

From design and performance simulations, we predict the total efficiency of the spectrometer system to be 24\%. This estimate is the product of several individual efficiencies shown in Table 2. The primary factors limiting the spectrometer's total efficiency are the initial efficiency with which light couples from the telescope optics to the spectrometer circuit via the slot antenna, and the efficiency with which the light in the 2D diffractive parallel-plate waveguide region couples into the receivers (Figure \ref{fig:cutouts}e). The antenna return loss, angular response function, and truncation of the beam by the cold stop all contribute to the total slot antenna coupling efficiency, with loss into the sidelobes of the antenna being the dominant source of loss. For the diffractive region, the transmitting array elements' coupling efficiency, the receiver circle's feedhorn antenna aperture efficiency and coupling efficiency, the light reflected at the receiver array/interference region interface, and the receiver array spillover all impact the Rowland circle receiver array's coupling efficiency (as shown in Figure \ref{fig:rec_eff}). Other contributions to the spectrometer efficiency that result in less significant losses include the efficiencies of the order-choosing filter, the reference coupler, the MKID coupling, and the transmission line.

\subsection{Stray Light}
\label{sec:stray}

\begin{center}
   \begin{figure} [ht]
   \hspace{1.5cm}
   \includegraphics[height=9cm]{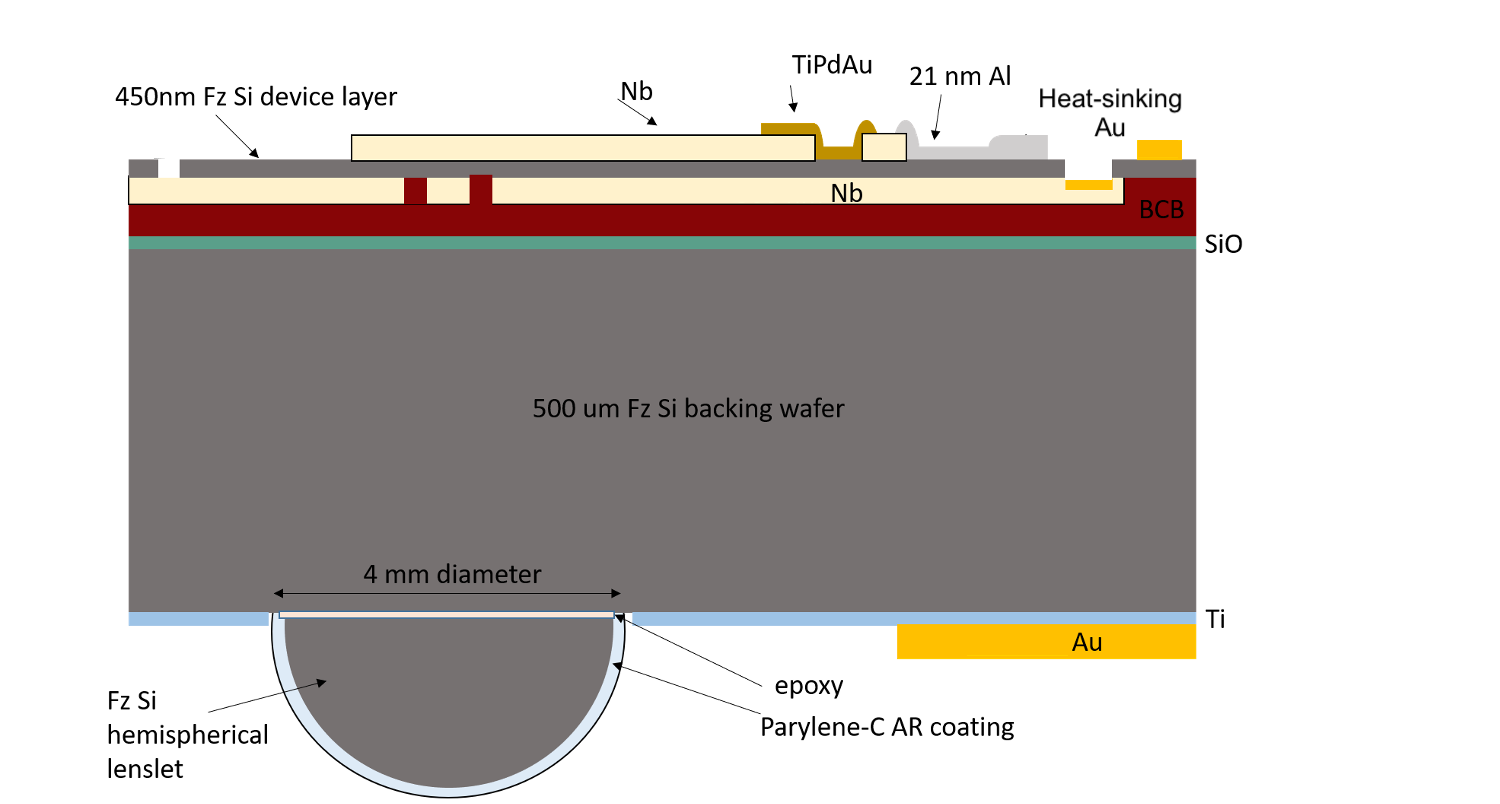}
   \caption[example] 
   {\label{fig:sideview} 
    A cartoon cross-section of the layers of the EXCLAIM spectrometer (not to scale).}
   \end{figure} 
\end{center}

Based on results from the $R{=}64$ $\mu$-Spec prototype, special design attention has been placed on predicting and mitigating the effects of stray light in the EXCLAIM spectrometers. One of the more significant design changes is the addition of a thin Ti coating layer (Figure \ref{fig:sideview}) with a thickness designed to target a sheet resistance optimized to terminate stray light in the Si backing wafer. Another significant change in design is the addition of a layer of normal-metal Au solely under the portion of the spectrometer housing the detectors, with a thickness designed to maximize the collecting area volume under the KIDs (Figure \ref{fig:sideview}) to serve as a trap for phonons excited by cosmic ray strikes.

A detailed accounting of stray light estimates and mitigation can be found in Table \ref{tbl:excel-table}. Other major stray light design considerations include an RF design that minimizes ground plane cuts, which are present in only the slot antenna aperture and the coupling capacitors and an Al feedline for the detectors' readout, which absorbs thermal radiation $>90$\,GHz incident from the coaxial cables from warmer temperature, along with powder filters at the inputs.

\begin{table}
  \caption{Stray-light design estimates and design elements for the R = 512 spectrometers. Details about the stray light control in the design of the telescope and receiver optics can also be found in Ref.~\cite{10.1117/12.2576254}.}
  \label{tbl:excel-table}
  \includegraphics[width=\linewidth]{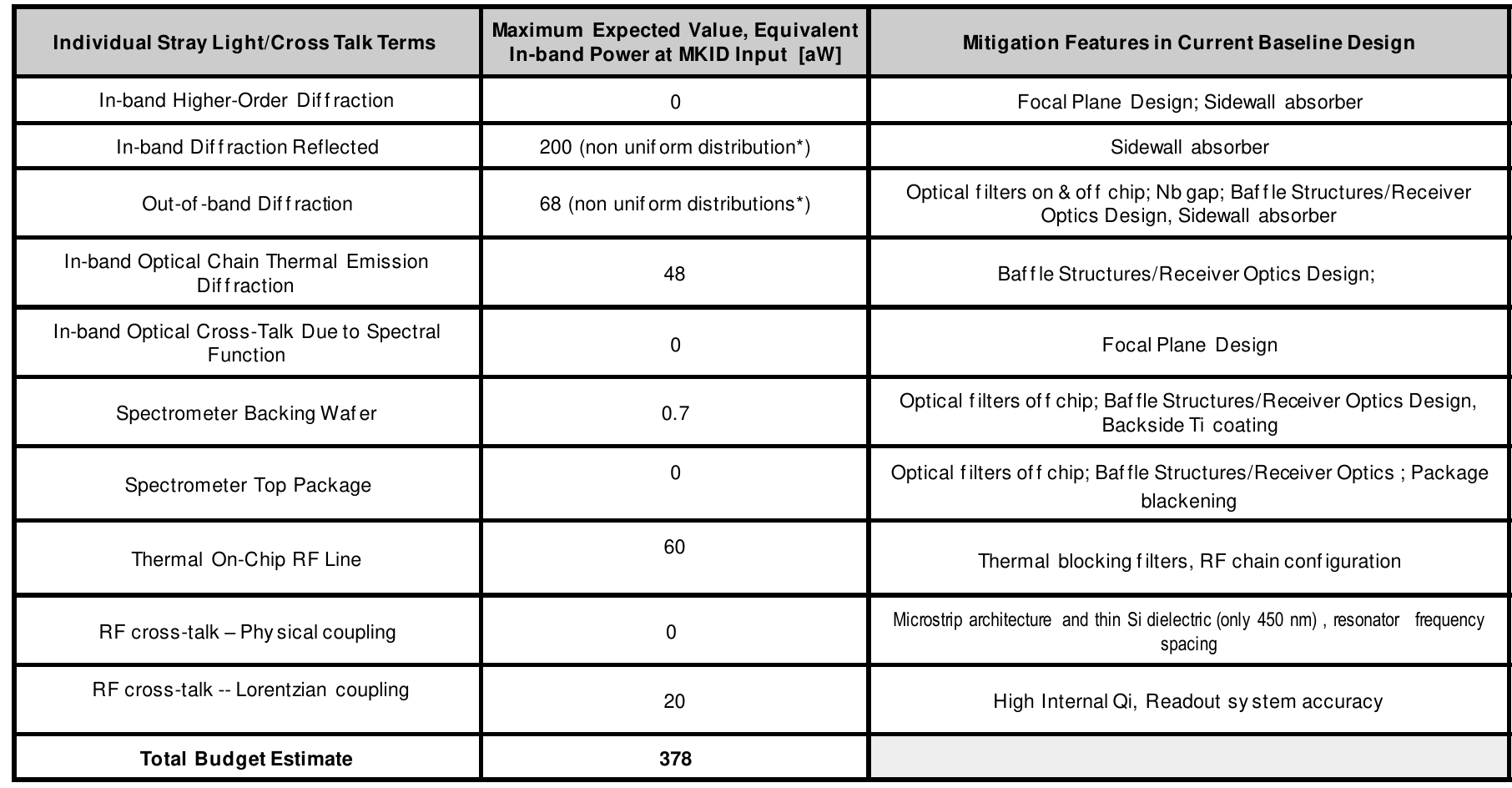}
\end{table}

\subsection{Fabrication}
\label{sec:fab}

The EXCLAIM $\mu$-Spectrometer fabrication heavily leverages the processes developed for the R = 64 prototypes \cite{patel2013fabrication,brown2015high}, with new steps based on prototype and test device results \cite{10.1117/12.2562446}. The $\rm \mu$-Spectrometer contains superconducting aluminum (Al) and niobium (Nb) films that are fabricated on two sides of the single-crystal silicon device layer of a silicon-on-insulator (SOI) wafer with a float zone (f-z) silicon device layer of 450-nm thickness. The fabrication process begins with patterning the niobium ground plane layer on an SOI wafer using a lift-off process to avoid roughening and etching through the silicon device layer.

The lift-off process contains a thin layer of germanium (Ge) hard mask, which is deposited on PMMA via sputtering or evaporation. Then, a thin layer of hexamethyldisilazane (HMDS) and photoresist is applied on top of germanium. A long-duration and low-temperature bake process is required to evaporate the solvent while the Ge is protected from thermal shock. A contact mask then patterns the resist. Next, Reactive Ion Etching (RIE) is used to etch germanium, and oxygen plasma is applied to remove the top layer of resist and to create an undercut necessary for the lift-off process. The silicon native oxide is then removed using a Buried Oxide Etch (BOE) prior to deposition of the niobium ground plane. After sputter-depositing the niobium, the lift off is completed in acetone.

The niobium ground plane side of the SOI wafer is then bonded to the front side of the f-z silicon wafer using benzocyclobutene (BCB) polymer. A combination of lapping and etching removes the handle layer of the SOI wafer. Then the buried oxide layer is etched with a BOE to expose the other side of the silicon device layer. Next, the top-side niobium is patterned with a similar lift-off process using a germanium hard mask. However, the lithography is performed in two steps, using a hard mask and through direct laser writing to define the sub-micron features in the slot antenna. A gold-palladium (AuPd) layer with a titanium (Ti) adhesion layer is then patterned via a lift-off process between and over areas of the niobium microstrip lines. A gold layer for heatsinking is also deposited and patterned via a lift-off process.

In the next step, the aluminum film is patterned to form the MKID array structure. To obtain a proper aluminum film profile, the native oxide on the silicon is removed by dilute hydrofluoric acid (HF dip) and a reverse bias clean prior to deposition. After sputter-depositing the aluminum, masked by the HMDS and photoresist, the aluminum is wet-etched in aluminum etchant and the remaining resists are stripped.

In the next step, niobium ground plane access vias are etched through the device layer for wirebonding purposes. Then, while the front of the wafer (containing the spectrometer circuit) is protected with a protective resist, the gold and titanium layers are patterned on the back of the supporting silicon backing wafer. The spectrometer chips are then defined by a deep reactive ion etch process. Finally, the final release of the chips from the wafer is performed in acetone. 


\section{Spectrometer Package}
\subsection{Design}

\begin{figure} [ht]
   \begin{center}
   \begin{tabular}{c} 
   \includegraphics[height=8cm]{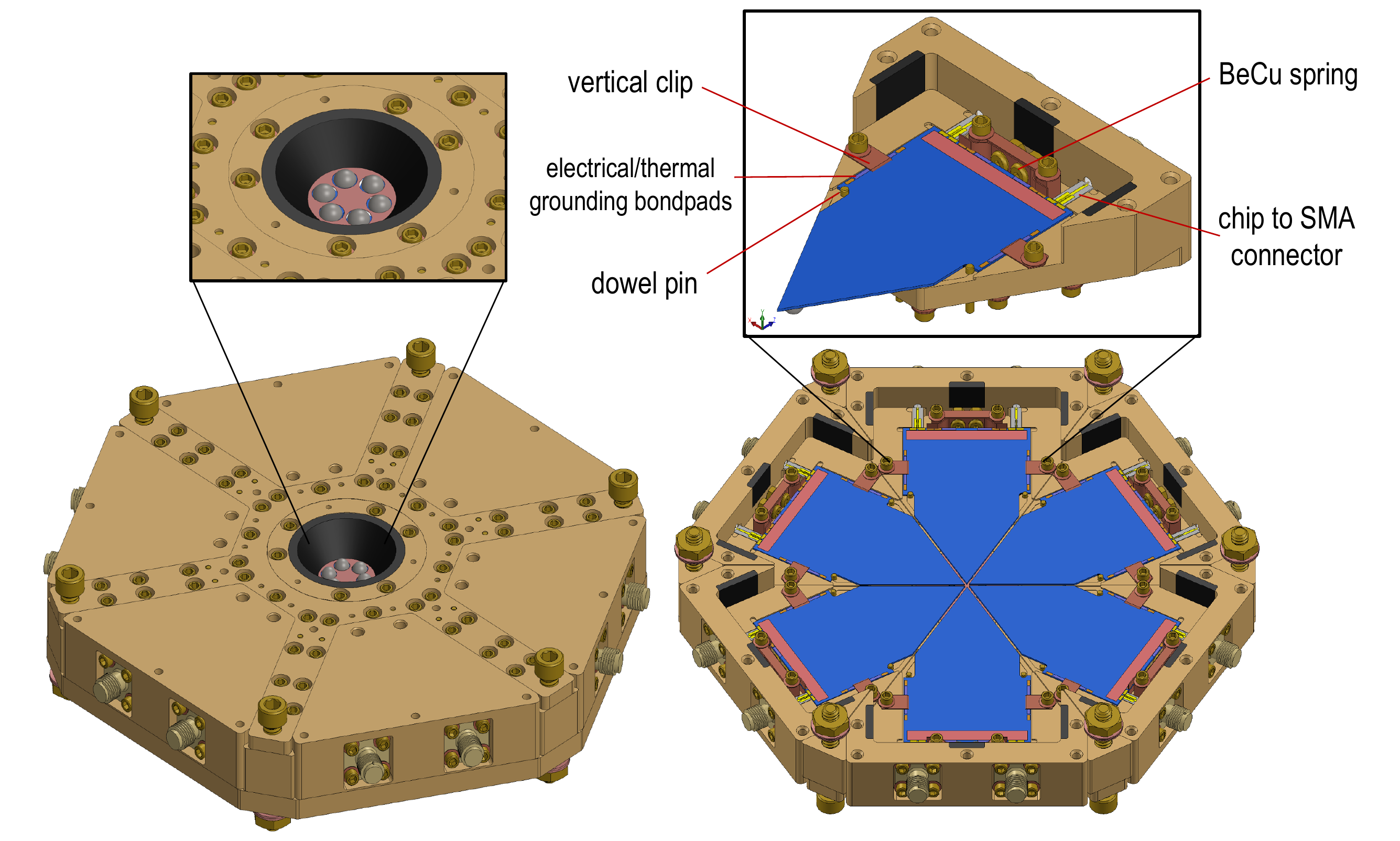}
   \end{tabular}
   \end{center}
   \caption[example] 
   { \label{fig:package} 
The six-spectrometer superstructure flight package. Left: The closed package, with SMA-connector connector pairs of the individual spectrometers visible along the side of the package, and a cutout featuring the six spectrometer lenslets in the package center. Right: A view of all six spectrometers mounted on the inside of the package, with a cutout showing the detailed components securing and aligning the individual spectrometer chips.}
   \end{figure}

EXCLAIM's flight-package is designed to house all six on-chip $\rm \mu$-Spectrometers in one superstructure (Figure \ref{fig:package}). The superstructure package is designed to allow individual spectrometers to be easily interchanged, so that the spectrometer chips with the highest performance and yield from any given wafer can be selected for the flight focal plane, and future on-chip spectrometers of different designs could also be inserted into the package if certain design elements are retained. This design approach also minimizes the re-packaging steps needed between the screening and characterization of individual spectrometers in our laboratory testbed and their integration into the flight focal plane package. The package is heat-sunk in flight through a copper bus to a continuous adiabatic demagnetization refrigerator (ADR) \cite{PIPERCADR}, which provides cooling to $100$\,mK. At these temperatures $10\times$ below the $T_c$ of the Al KIDs, noise contributions from thermal generation are negligible, and the spectrometers are insensitive to the ADR temperature variations. 

The focal plane of the telescope's cold optics illuminates the AR-coated lenslets of the six spectrometers, with an insert covered in light-absorbent blackening epoxy mitigating side reflections. In each segment of the superstructure, a spectrometer is held in place with a series of clips, pins, and springs. Two vertical clips, situated along the two long walls of the spectrometer's interference region, constrain the chip vertically. Along the horizontal plane of the spectrometer chip, a BeCu spring situated on the detector-end gently pushes the chip to rest against two alignment pins situated near the vertical clips on both sides of the spectrometer (Figure \ref{fig:package}). A V-groove fixes the axis of rotation and a pin on a flat feature locks the clocking in a kinematic arrangement. The BeCu spring allows for thermal contraction of the package without stressing the spectrometer chip. The registration of the focal plane to the silicon focusing lens incorporates the coefficient of thermal expansion of the optics tubes and suspensions, and a jig controls the tolerance of the focal plane placement during assembly. An equal-length re-entrant carbon fiber suspension design cancels differential expansion of tube components \cite{EXCLAIMjatis}.

Each spectrometer segment contains two coplanar-waveguide (CPW) to microstrip transition microwave fanout boards at the input and output of the spectrometer's detector array feedline. These fanout boards feature a coplanar-waveguide (CPW) to microstrip transition and connect via wirebonds to the SMA pins and to a CPW bondpad on the spectrometer chip, which transitions to the on-chip microstrip MKID feedline. It is through these two SMA connectors that each spectrometer's detector array is biased and readout. The bottom cover of the superstructure package is covered in pockets of light-absorbent blackening epoxy to mitigate reflections and stray light within the package.

\section{Conclusion: Status}
The design phase for EXCLAIM's spectrometers is complete, and the first spectrometers are expected to be yielded in early fall of 2022 to begin to undergo characterization testing.

\begin{figure} [h]
   \begin{center}
   \begin{tabular}{c} 
   \includegraphics[height=9cm]{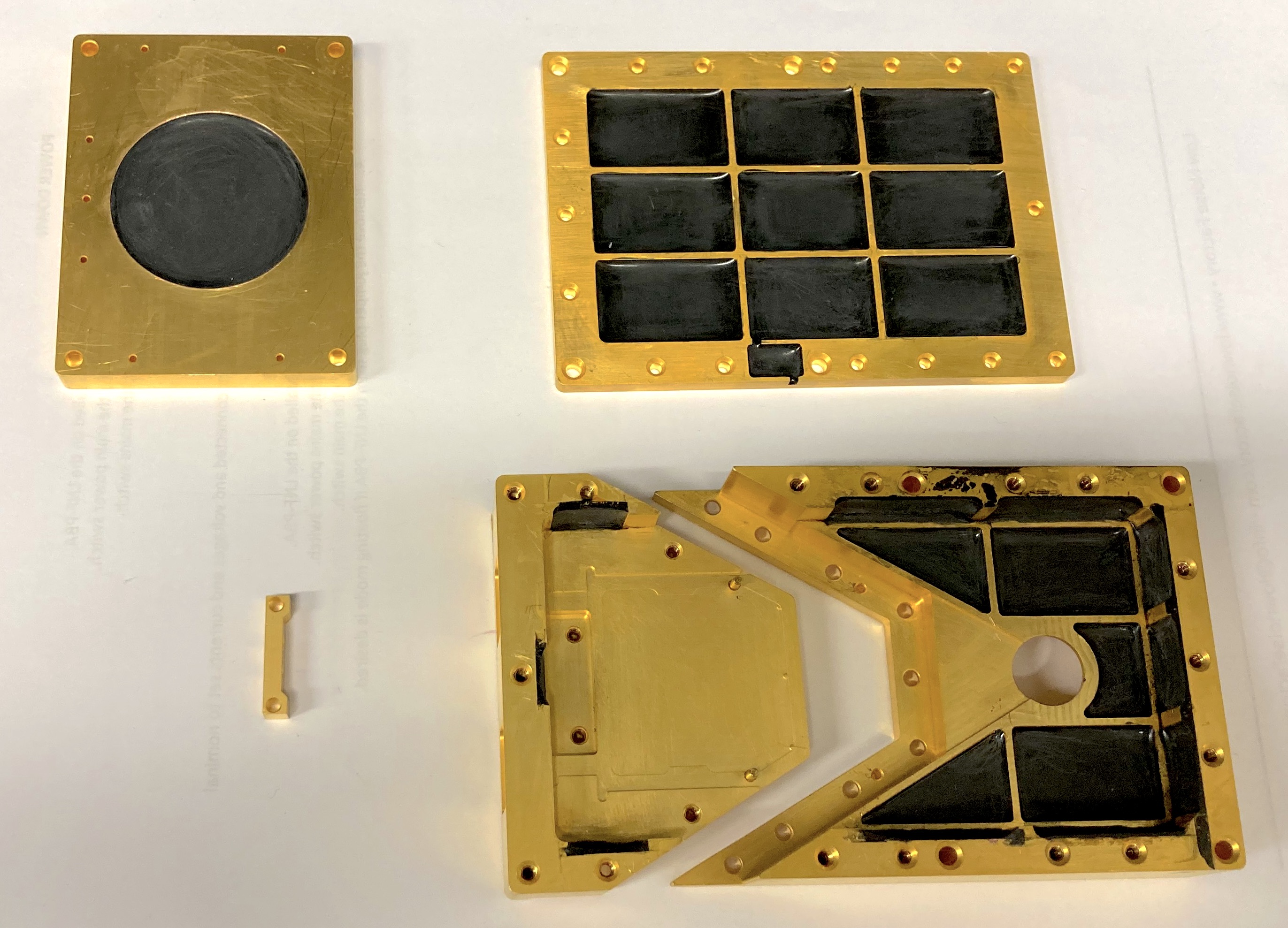}
   \end{tabular}
   \end{center}
   \caption[example] 
   { \label{fig:photo} 
A photograph of the disassembled single spectrometer package used to characterize the individual spectrometers. The spectrometer is mounted in the two-part element in the bottom right, with the lenslet oriented to be visible through the circular hole. On the top right is the package lid, and on the top left is a removable cover for the spectrometer lenslet. On the bottom left is a small bracket mount for the springs that are used to align the spectrometer chip in the package.}
   \end{figure}

The EXCLAIM six-spectrometer superstructure package design is complete. The spectrometer characterization testbed is configured to test individual spectrometers, so an additional package has been designed to house a single, easily interchangeable chip. This single-chip package has been completed (Figure \ref{fig:photo}), and has been used to develop processes for creating and applying the blackening epoxy. As the EXCLAIM team yields spectrometers, they will be characterized individually in the testbed using this package, and the six best-performing spectrometers will be selected to include in the final flight package.

\acknowledgments 
 
This material is based upon work supported by NASA under award number 80GSFC21M0002.
\newpage
\bibliography{report} 
\bibliographystyle{spiebib} 

\end{document}